\documentclass[useAMS,usenatbib]{mn2e}

\usepackage{lscape}
\usepackage{tabularx}
\usepackage{xtab}
\usepackage{xspace}
\usepackage{graphicx}
\usepackage{placeins}
\usepackage{amsfonts,amsmath,amssymb}
\usepackage{url}
\usepackage[usenames,dvipsnames]{color}
\usepackage{times}
\usepackage{soul}

\usepackage{enumerate}
\def\aj{AJ}%
\def\araa{ARA\&A}%
\def\apj{ApJ}%
\def\apjl{ApJ}%
\def\apjs{ApJS}%
\def\apss{Ap\&SS}%
\def\aap{A\&A}%
\def\aapr{A\&A~Rev.}%
\def\aaps{A\&AS}%
\def\mnras{MNRAS}%
\def\pasp{PASP}%
\def\nat{Nature}%
\def\aplett{Astrophys.~Lett.}%
\def\physrep{Phys.~Rep.}%
\newcommand{\msun}{~\mathrm{M}_{\odot}}

\title[Unorthodox evolution of major merger remnants]{The unorthodox evolution of major merger remnants into star-forming spiral galaxies}

\author[Sparre \& Springel]{\parbox[t]{\textwidth}{
		Martin Sparre$^{1,2}$\thanks{Sapere Aude Fellow, E-mail:sparre@dark-cosmology.dk} and
		Volker Springel$^{1,3}$
		\vspace*{6pt}} \\
$^1$Heidelberger Institut f{\"u}r Theoretische Studien, Schloss-Wolfsbrunnenweg 35, 69118 Heidelberg, Germany\\
$^2$Dark Cosmology Centre, Niels Bohr Institute, University of Copenhagen, Juliane Maries Vej 30, 2100 Copenhagen, Denmark\\
$^3$Zentrum f\"ur Astronomie der Universit\"at Heidelberg, Astronomisches Recheninstitut, M\"onchhofstrasse 12-14, 69120 Heidelberg, Germany\\
}

\begin{document}

\date{\today}

\pagerange{\pageref{firstpage}--\pageref{lastpage}} \pubyear{2015}
\maketitle

\label{firstpage}

\begin{abstract}
Galaxy mergers are believed to play a key role in transforming star-forming disk galaxies into quenched ellipticals. Most of our theoretical knowledge about such morphological transformations does, however, rely on idealised simulations where processes such as cooling of hot halo gas into the disk and gas accretion in the post-merger phase are not treated in a self-consistent cosmological fashion. In this paper we study the morphological evolution of the stellar components of four major mergers occurring at $z=0.5$ in cosmological hydrodynamical zoom-simulations. In all simulations the merger reduces the disk mass-fraction, but all galaxies simulated at our highest resolution regrow a significant disk by $z=0$ (with a disk fraction larger than 24\%). For runs with our default physics model, which includes galactic winds from star formation and black hole feedback, none of the merger remnants are quenched, but in a set of simulations with stronger black hole feedback we find that major mergers can indeed quench galaxies. We conclude that major merger remnants commonly evolve into star-forming disk galaxies, unless sufficiently strong AGN feedback assists in the quenching of the remnant.
\end{abstract}

\section{Introduction}

Traditionally, the visual appearance of galaxies has motivated dividing them into irregulars, spirals and ellipticals \citep{1926ApJ....64..321H}. An important difference between spirals and ellipticals is that the former are star-forming, whereas the latter are more likely quenched \citep{1998ARA&A..36..189K}. These classifications are also in line with the presence of the Tully-Fisher relation \citep{1977A&A....54..661T}, which describes the relation between stellar luminosity and rotation velocity for spiral galaxies, and the Faber--Jackson relation \citep{1976ApJ...204..668F}, which encodes the scaling of luminosity with the velocity dispersion of ellipticals. In-between spirals and ellipticals are the lenticular galaxies, which are essentially quenched galaxies with a dominating spherical component and an old stellar disk. Lenticular and spiral galaxies have a different normalisation in their Tully--Fisher relations, with the amount of stellar rotation in the spirals being larger than for lenticulars.

In the modern view of galaxy evolution, the transformation from disks to ellipticals is discussed in terms of the so-called \emph{star formation main sequence}, which is a relation between a galaxy's star formation rate (SFR) and stellar mass ($M_*$). Numerous observational studies have established this relation \citep{2007ApJ...660L..43N,2007ApJS..173..267S,2011A&A...533A.119E,2014ApJS..214...15S,2015ApJ...801...80L,2016ApJ...817..118T,2016ApJ...820L...1K}, which is a power-law with an observed scatter around 0.2--0.3 dex. The normalisation of the relation is declining with time, implying that galaxies in the early Universe were typically forming stars more rapidly than at the present day.

But only a subset of the galaxies follow this main sequence relation. \emph{Quenched} galaxies, for example, have SFRs that are significantly lower than predicted by the \emph{main sequence}. On the other hand, there are also starbursts that transform their interstellar gas into stars on unusually short timescales. These galaxies typically have much larger SFRs than predicted by the main sequence \citep{2011ApJ...739L..40R,2012ApJ...747L..31S}. Note, however, that the definition of a starburst is somehow ambiguous, since a relatively gas-poor galaxy can in principle consume its gas on a short timescale (if it is in a \emph{bursty mode}), but it does not guarantee that the galaxy is more star-forming than a normal main sequence system of similar mass, simply because of the small absolute amount of gas available for star formation. The galaxies with a much larger SFR than predicted by the main sequence are therefore only a subset of all the galaxies consuming their gas in a bursty mode (for further discussions, see \citealt{2009ApJ...698.1437K} and \citealt{2013seg..book..491S}). In this paper we refer to galaxies with a SFR well above what is predicted by the main sequence as starbursts\footnote{In our companion paper, \citealt{2016arXiv160408205S}, we studied the gas consumption timescales of major mergers, where we hence used a different starburst definition.}.

Galaxies following the star formation main sequence are most often disk galaxies, whereas the starbursts are often interacting systems \citep{2013ApJ...778..129H}. The quenched galaxies are more likely to be elliptical galaxies \citep{2011ApJ...742...96W}. In the most na{\" i}ve interpretation of the main sequence in terms of an evolutionary scenario, galaxies form the majority of their stars while evolving on the main sequence as spiral galaxies. Starbursts occur when galaxies merge, and merger-induced feedback processes afterwards quench the merger remnant, leaving behind an elliptical galaxy. However, it is not clear whether the feedback-induced quenching is really that strong, or whether it is lasts for a significant time. Otherwise, a new disk may promptly grow.

One of the conjectured processes for the quenching and morphological transformation of galaxies is feedback caused by the black holes in the center of galaxies \citep{1998A&A...331L...1S,2003ApJ...596L..27K,2005ApJ...618..569M,2006MNRAS.373L..16F,2009Natur.460..213C}. Due to gravitational tidal torqueing, the rate of gas infall into galactic centres is especially high during galaxy mergers, suggesting that this should be conducive to rapid black hole growth. Indeed,
 simulations that assume that some fraction of the accretion energy couples to the gas predict the creation of strong outflows of gas, which can quench galaxies \citep{2005MNRAS.361..776S,2008ApJS..175..356H}. Observationally, AGN (Active Galactic Nucleus) activity and quenching seem to be related, although the picture is somewhat muddled and seems more complicated than expected for the simplest main sequence evolution scenario. For example, most AGN host galaxies show no signs of recent mergers \citep{2011ApJ...726...57C,2011ApJ...727L..31S,2012ApJ...744..148K,2013A&A...549A..46B,2014MNRAS.439.3342V}. There is, however, still observational support for a connection between merger-induced starbursts and AGN activity, since galaxy mergers show evidence for an increased AGN fraction compared to more isolated galaxies \citep{2007AJ....134..527W,2010ApJ...716L.125K,2013MNRAS.430..638S,2013MNRAS.435.3627E,2014MNRAS.441.1297S}.

Considering that quenched galaxies are observed to be mainly ellipticals and that many theoretical models predict galaxies to quench by going though a merger-induced starburst (causing AGN feedback), it is important to study the morphological transformation of galaxies in simulations of mergers. Idealised simulations have shown that it is possible for galaxy disks to survive a major merger \citep{2005ApJ...622L...9S}. \citet{2006ApJ...645..986R} also showed that merger remnants can potentially have star-forming disks provided that there is enough gas available at the time of final coalescence of the galaxies. Furthermore, \citet{2009MNRAS.398..312G} simulated a major merger in a full cosmological setup and identified a dominating stellar disk in the merger remnant. These simulation results challenge the simplest possible picture, where major mergers always destroy disks and produce spheroidal galaxies that are quenched by the associated black hole feedback. In a cosmological simulation, \citet{2003ApJ...597...21A} also identified a galaxy with a thin disk mostly formed in-situ after a merger event, and also a thicker disk consisting of stars accreted from satellites with orbital planes coincident with the disk plane. A post-merger galaxy can thus be \emph{disky} if the disk survives the merger, or because there is sufficient gas left to reform a disk. Finally, mass can also be added to the disk through the accretion of satellites.

\begin{figure*}
\centering
\includegraphics[width = 0.98 \textwidth]{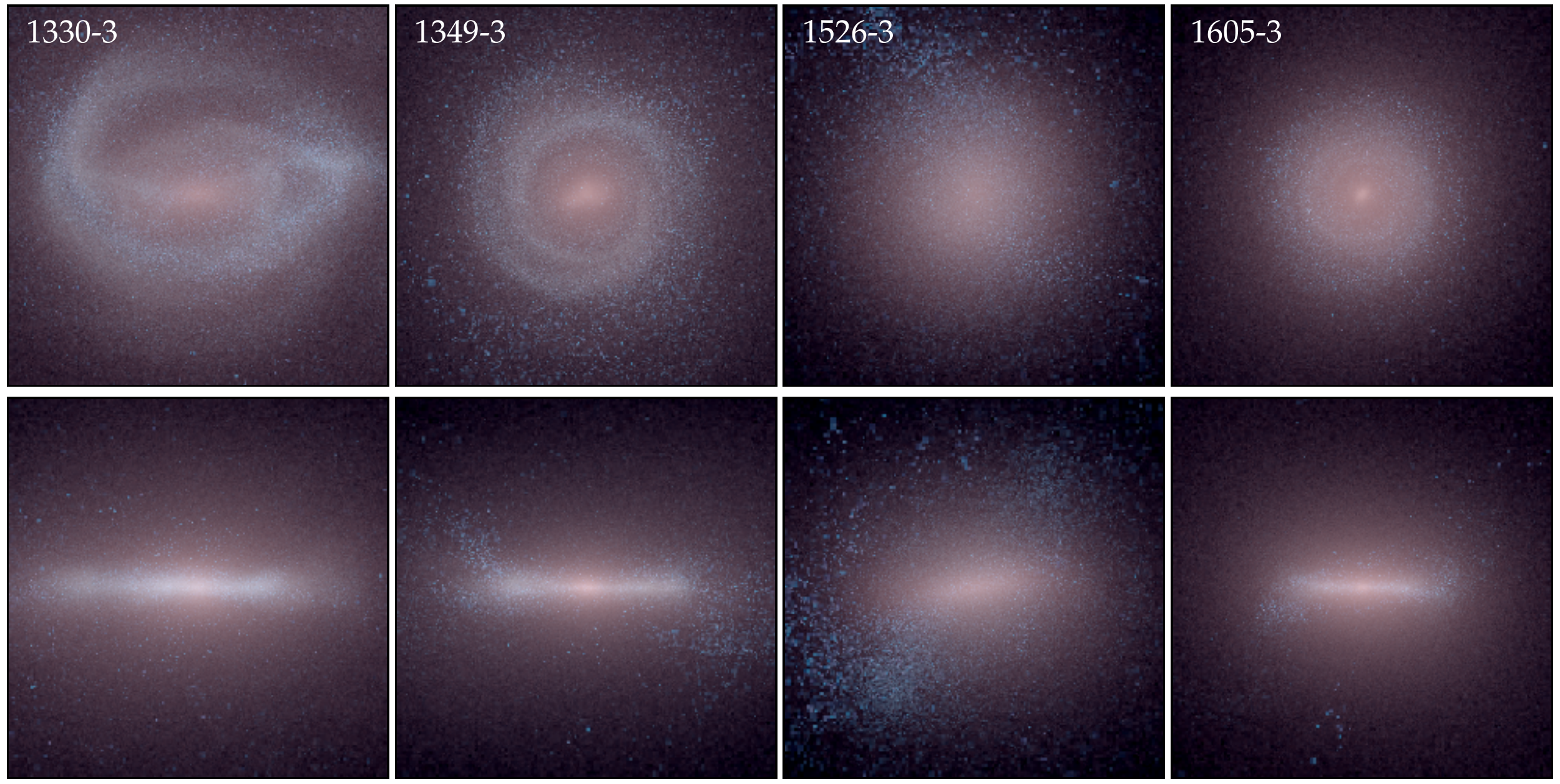}
\caption{The distribution of stars at $z=0$ for the merger remnants from the four high-resolution simulations. \emph{Upper panels} show face-on projections, and the \emph{lower panels} show the edge-on versions. The \emph{blue}, \emph{green} and \emph{red} colour are determined based on the luminosity in the $U$, $B$ and $K$ band, respectively. The $z=0$ morphology of a galaxy that experienced a  major merger at $z=0.5-1$ can take many shapes: the 1349-3 and 1605-3 simulations have a disk with many young (\emph{blue}) stars. Furthermore, spiral arms are visible in the face-on projection of 1349-3. In addition to showing a disk 1330-3 also has a bar. A disk is not clearly revealed in the image of 1526-3 (the off-center distribution of recently formed blue stars are caused by an accreted star-forming galaxy). The width of each panel is 40 kpc.}
\label{PlotDensityMapAtRedshift0}
\end{figure*}

One way to study galaxy evolution scenarios is through large-scale cosmological simulations, such as \emph{Illustris} \citep{2013MNRAS.436.3031V,2014MNRAS.445..175G}, EAGLE \citep{2015MNRAS.446..521S,2015MNRAS.450.1937C}, MassiveBlack-II \citep{2015MNRAS.450.1349K} and Horizon-AGN \citep{2016arXiv160201941V,2016arXiv160509379K}. Simulations of this kind have made it possible to study a diverse set of phenomena, for example the formation and evolution of massive compact galaxies at $z=2$ \citep{2015MNRAS.449..361W,2015arXiv151005645F,2016MNRAS.456.1030W}, the bimodial surface brightness distributions at $z=0$ \citep{2015MNRAS.454.1886S}, or the merger rate of galaxies \citep{2015MNRAS.449...49R}. A limitation of these simulations is that they do not have the resolution to accurately capture starbursting gas that appears in galaxies \citep{2015MNRAS.447.3548S}. For resolving such starbursting gas in merger-induced starbursts (for the galaxy formation model of Illustris) a $\simeq 10-40$ times better mass resolution
 is necessary \citep{2016arXiv160408205S}. Other numerical implementations of star formation may require an even higher resolution to fully resolve the star-bursting gas \citep{2014MNRAS.442L..33R}.

To reliably model the physical processes in major mergers it is therefore necessary to perform simulations with increased resolution compared to Illustris.  In this paper we study the transformation of stellar morphology with a suite of major merger simulations in a zoom setup where the resolution is indeed much higher, allowing in particular a better representation of the starburst regions (and hence also of the black hole feedback). These simulations are fully cosmological, which means that the circumgalactic gas and processes such as gas-fueling in the post-merger stage are included in a self-consistent way, as prescribed by the $\Lambda$CDM paradigm. Our suite of major mergers is therefore ideal to test whether the simulations are consistent with the observed galaxy evolution scenario, and to make new predictions for galaxies in the real universe.

The aim of the paper is to analyse the detailed stellar morphology of cosmological merger remnants, and to check whether these are indeed consistent with quenched ellipticals (which would be in line with simple interpretations of the \emph{main sequence evolution scenario}), or whether some of them remain spiral galaxies, which would also not be too surprising given some previous  simulations found this unexpected outcome  \citep{2005ApJ...622L...9S,2009MNRAS.398..312G,2009ApJ...691.1168H}.
We note that while semi-analytic models of galaxy formation have generally assumed that major merger remnants destroy disks and produce a spheroid, they have also allowed for a regrowth of a disk with time depending on the amount of gas left in the remnant \citep[e.g.][]{1993MNRAS.264..201K}.

Or study is structured as follows. We first introduce our simulations in Section~\ref{methods}. In Section~\ref{MorphologySection}, we study both the $z=0$ morphology of the merger remnants, and the evolution of the stellar disks. We will among other things see how major mergers affect the stellar morphology by comparing merger simulations with galaxies of similar mass but with a more quiescent evolutionary history \citep[from the Auriga simulation suite;][]{2016arXiv161001159G}. In Section~\ref{StrongAGN}, we perform additional simulations with stronger AGN feedback to study how this affects the merger remnants. Finally, we give a discussion of our results in Section~\ref{Discussion} and present our conclusions in Section~\ref{Conclusion}.

\section{The merger simulations} \label{methods}

In this paper we study the same major merger simulations previously discussed in \citet{2016arXiv160408205S}. The sample contains four different mergers occurring at $0.5<z<1$, with stellar mass ratios of the merging galaxies ranging from 1.00 to 1.51. The $z=0$ stellar masses are in the range $10^{10.61}<M_*/\msun<10^{11.04}$, and the halos have virial masses $10^{12.00}<M_{200}/\msun<10^{12.27}$.

These galaxies were selected by analysing the merger trees of Illustris \citep{2015MNRAS.449...49R}. For an initial selection we picked out the $z=0$ galaxies with merger trees obeying the following criteria:
\begin{itemize}
\item The main $z=0$ galaxy has a stellar mass of $10^{10.5}< M _*/\msun <10^{11.0} $.
\item At $z=1$ exactly two galaxies with a stellar mass ratio in the range, $0.80<\mu<1.25$, are present in the merger tree.
\item At $z=0.2$ and $ z=0.5$ only one massive galaxy is present in the merger tree, i.e. if $M_1$ is the stellar mass of the main progenitor, then no other galaxy from the merger tree has $M_2>M_1 / 3$ at any of these two redshifts. This criterion makes sure that the galaxy has time to relax until $z=0$.
\item The $z=0$ dark matter mass bound to the central galaxy should be larger than half of the $M_{200}$-value of the halo in which the galaxy resides. This ensures that the $z=0$ galaxy is dominating the halo, and that the galaxy is isolated. 
\end{itemize}
With these selection criteria we obtain candidates that undergo a major merger between $z=1$ and $z=0.5$ and furthermore evolve relatively isolated between $z=0.5$ and $z=0$. Furthermore, the $z=0$ stellar masses and halo masses are comparable to that of the Galaxy. These selection criteria yielded a total of 14 merger candidates from the Illustris simulation box. We randomly selected four of these 14 galaxies to arrive at our current sample of mergers.

Each of the four merger systems was simulated at three different resolutions, with `zoom factors' of 1, 2 and 3, corresponding to mass resolutions factors of 1.4, 11.4 and 38.5 times finer than in the Illustris simulation. The maximum physical softening is 0.64, 0.32 and 0.21 kpc, respectively. To carry out the simulations we used a zoom-in technique, where the spatial resolution is high in the vicinity of the galaxy of interest and progressively lower further away. This makes it possible to carry out fully self-consistent cosmological simulations of individual galaxies at a small fraction of the computational cost of the Illustris simulation.

The galaxy formation model, which is closely based on \citet{2014MNRAS.437.1750M} and \citet{2013MNRAS.436.3031V}, is the same as used for the Auriga simulation project \citep{2016arXiv161001159G}. The hydrodynamical equations are solved with the moving-mesh approach used by the AREPO code \citep{2010MNRAS.401..791S}, which ensures an accurate treatment of shocks and fluid instabilities in cosmological environments as well as low advection errors \citep{2012MNRAS.423.2558B,2012MNRAS.424.2999S,2012MNRAS.425.2027K,2012MNRAS.425.3024V,2012MNRAS.427.2224T,2013MNRAS.429.3341B,2013MNRAS.429.3353N,2014MNRAS.442.1992H}. For details about the merger simulations and the galaxy formation model, we refer to \citet{2016arXiv160408205S}.

Merger trees were constructed by considering the stellar population particles contained in the two most massive halos at $z=0.93$, which is before the major merger occurs in all simulations. At each snapshot, we determine the galaxies which have most stars in common with each of the selected galaxies at $z=0.93$. With this method, we track two progenitor branches before the merger. After the merger, the two $z=0.93$ progenitors have the same descendant galaxy.

\subsection{Setup for runs with stronger AGN feedback}\label{SetupStrongAGN}

For each of our simulations with a `zoom factor' of 2 we have performed additional simulations where the AGN feedback is gradually increased. In our AGN feedback model, which relies on the model of \citet{2000MNRAS.311..346N}, the black hole accretion rate is inversely proportional to the cooling function, $\Lambda (Z,T)$, of the gas in a galaxy. The temperature $T$ is here set to the virial temperature of a halo, and the metallicity, $Z$, is a model parameter. Decreasing the metallicity will increase the strength of the black hole accretion rate and also make AGN feedback stronger. In our set of simulations with stronger AGN feedback we set the metallicity in the cooling function to 1.0, 0.4, 0.2 and 0.1 $\text{Z}_{\sun}$. We refer to these runs as having \emph{normal}, \emph{semi-strong}, \emph{strong} and \emph{very strong} AGN feedback, respectively.

\section{Stellar morphology of the merger remnants} \label{MorphologySection}

Morphological classification of galaxies can be done in several ways. Traditionally, galaxy types have been distinguished based on their visual appearance \citep{1926ApJ....64..321H}. A more rigid morphological characterization method, which is especially well suited for observational applications, is based on surface brightness fitting \citep[described in e.g.][]{2002AJ....124..266P,2010AJ....139.2097P}. In galaxy simulations, where information for the coordinates and velocities are known, a disk--bulge decomposition can usually be done based on the angular momentum of each stellar population particle \citep{2009MNRAS.396..696S}. In the following, we classify the simulated galaxies according to each of these methods, and study how the disk evolution connects to the colours and quenching properties of the galaxies.

\begin{figure*}
\centering
\includegraphics[width = 0.98 \textwidth]{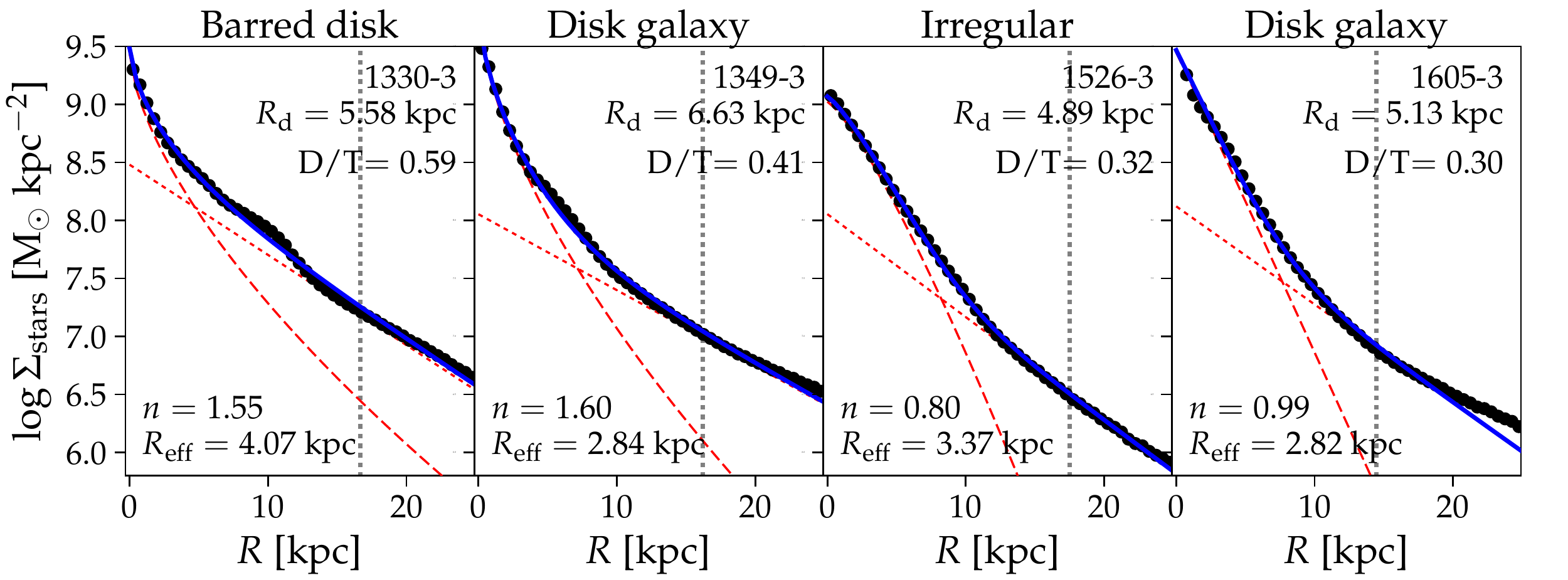}
\caption{The stellar mass surface density as a function of radius for each of the high-resolution galaxies at $z=0$ (\emph{black circles}). We measure this in a face-on projection. The \emph{thick blue line} shows a model containing a bulge and a disk component (their contributions are shown as \emph{dashed} and \emph{dotted red lines}, respectively). The disk fractions, D$/$T, are between 30 and 59\%, and the disk scale radii are 5--7 kpc. The \emph{dotted vertical lines} indicate a radius of $0.1\times R_{200}$. Only stars inside $0.15\times R_{200}$ are included in the fits. Above each panel the classification revealed by the visual classification in Figure~\ref{PlotDensityMapAtRedshift0} is noted.}
\label{PlotStellarSurfaceBrightness}

\centering
\includegraphics[width = 0.98 \textwidth]{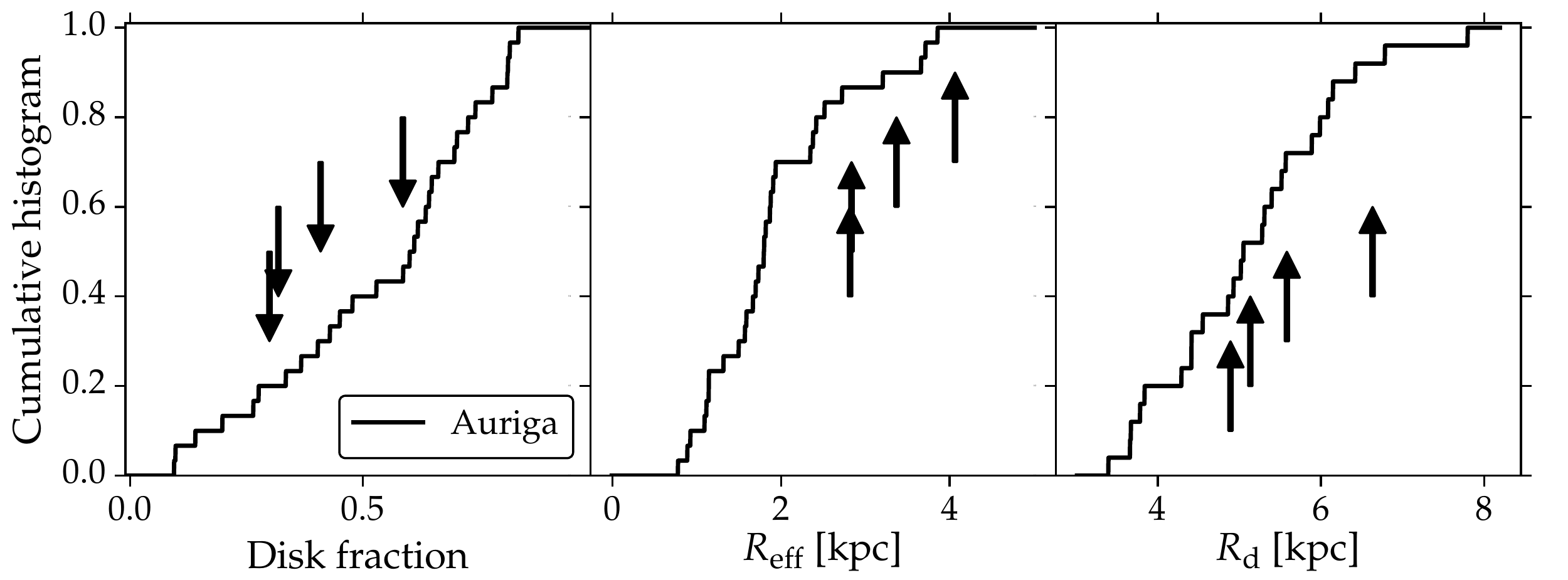}
\caption{A comparison of $z=0$ properties between our merger remnants (each high-resolution simulation is marked with an arrow) and the 30 galaxies from the Auriga simulation suite (shown as cumulative histograms). We compare the disk fraction, the effective radius of the bulge and the disk scale radius. The four arrows in each panel show the values obtained from the fit in Figure~\ref{PlotStellarSurfaceBrightness}. The merger remnants have larger effective bulge radii than the median of the Auriga sample. The disk fractions and disk scale ratio on the other hand are well overlapping with the distribution from the Auriga galaxies.}
\label{GG_comparison}

\end{figure*}

\subsection{Visual inspection of the merger remnants}

As a first step in studying the stellar morphology we create a composite $U$-, $B$- and $K$-band image of the stellar light distribution of the stars in our $z=0$ merger remnants, see Figure~\ref{PlotDensityMapAtRedshift0}. This allows us to visually classify the morphology of the galaxies. We show the galaxies in a face-on (\emph{upper panels}) and an edge-on projection (\emph{lower panels}) for each of our high-resolution simulations with a `zoom factor' equal to 3. The projections are determined based on the moment-of-inertia tensor of the stars in a galaxy. The figure reveals that two of our four major merger remnants (1349-3 and 1605-3) have a well-established disk containing young, recently formed stars. The edge-on projection of 1349-3 also reveals the presence of spiral arms, implying that spiral structure is not only present in galaxies with a quiescent merger history (as for the simulation studied in \citealt{2016arXiv160401027G}). For the 1330-3 galaxy, the edge-on projection also reveals a thin disk, and the head-on projection furthermore reveals a bar. For 1526-3, the images show no signs of a star-forming disk, but rather a red spherical component surrounded by some recently formed (\emph{blue}) stars with an irregular distribution. Based on the images we conclude that our major merger remnants fall into the category of either disk, bar or irregular galaxies.

\subsection{Surface density profiles}\label{SurfaceDensityProfiles}

The above considerations are made entirely based on images showing the stellar light distributions in different bands. An observationally motivated quantification of the morphology for the $z=0$ merger remnants is presented in Figure~\ref{PlotStellarSurfaceBrightness}, which shows the stellar mass surface density (\emph{black circles}) in a face-on projection of the inner 25 kpc of the galaxies. Each profile is modeled with a contribution from a S\'ersic profile (\emph{red dashed line}) and an exponentially decaying profile (\emph{red dotted line}). These components describe the contribution from the bulge and the disk, respectively. The sum of the two components is shown by the \emph{thick blue line}. We have simultaneously fitted the parameters describing these profiles, and based on the stellar mass in each component we derive the mass fraction of stars in the disk (the `disk fraction', or simply just D$/$T), which is listed in each panel. Also listed is the disk scale length ($R_\text{d}$), where the surface density of the disk declines by a factor of $e$. When performing the fits we only include stars within\footnote{Our standard choice in this paper is to analyse stars within 10\% of $R_{200}$ when we determine disk properties. When fitting surface brightness profiles we, however, allow for an exception and include stars within 15\% of $R_{200}$. This is done to better fit the scale radius of the disk, which is most easily determined in the radius-range of 8-15\% $R_{200}$. If we only included stars within 10\% $R_{200}$ we would get unreasonably steep disk profiles, and also too high disk fractions in our modelling.} 15\% of $R_{200}$.

The fits yield disk fractions of $30\%-59\%$. The spherically symmetric bulge component dominates the inner 5 kpc of all the galaxies, and at larger radii the disk is dominant. These galaxies thus have more massive and more radially distributed bulges than the Milky Way. The disk scale radii ($R_\text{d}$) are 5--6 kpc, which is $\simeq$3 times larger than observed for the Milky Way ($2.15\pm 0.14$ kpc; \citealt{2013ApJ...779..115B}). Based on these characteristics the merger remnants differ from classical disk galaxies due to their dominating bulges. It is, however, interesting that these merger remnants still exhibit visible disks in the images (Figure~\ref{PlotDensityMapAtRedshift0}) and in the surface density profiles (Figure~\ref{PlotStellarSurfaceBrightness}).

\citet{2009MNRAS.398..312G} also studied the evolution of the surface brightness profile of galaxies undergoing a cosmological merger.
Their merger remnant ends up with a $z=0$ disk dominating over the bulge\footnote{For various definitions they find disk fractions in the range, 0.53--0.67.}. We are therefore not the first to suggest that remnants of major mergers can have significant disks. Their disk fraction is as high as we find for 1330-3 (the bar galaxy). We regard our merger remnants to have disk fractions consistent with the galaxy from \citet{2009MNRAS.398..312G}. We note that the physical setup of our simulations is very different, however.

The best fit parameters from our surface brightness modeling of our $z=0$ merger remnants are compared to the 30 galaxies of the Auriga simulation in Figure~\ref{GG_comparison}. The Auriga simulation suite uses the exact same galaxy formation model as in our merger simulations, and the only physical difference is the initial conditions. The Auriga galaxies have $z=0$ halo masses of around $M_{200}\approx 10^{12}\msun$, and the haloes are selected to be isolated at $z=0$. No further constraints are enforced in their initial condition selection. The galaxies show a huge variety of mass accretion histories with some galaxies undergoing major mergers and other halos having a more quiescent evolution. The average properties of the galaxies from Auriga therefore represents what is expected for \emph{normal star-forming galaxies}. In \citet{2016arXiv161001159G} it was indeed also found that the galaxies from the Auriga simulation had stellar disk profiles and rotation curves consistent with observations.

\begin{figure*}
\centering
\includegraphics[width = 0.98 \textwidth]{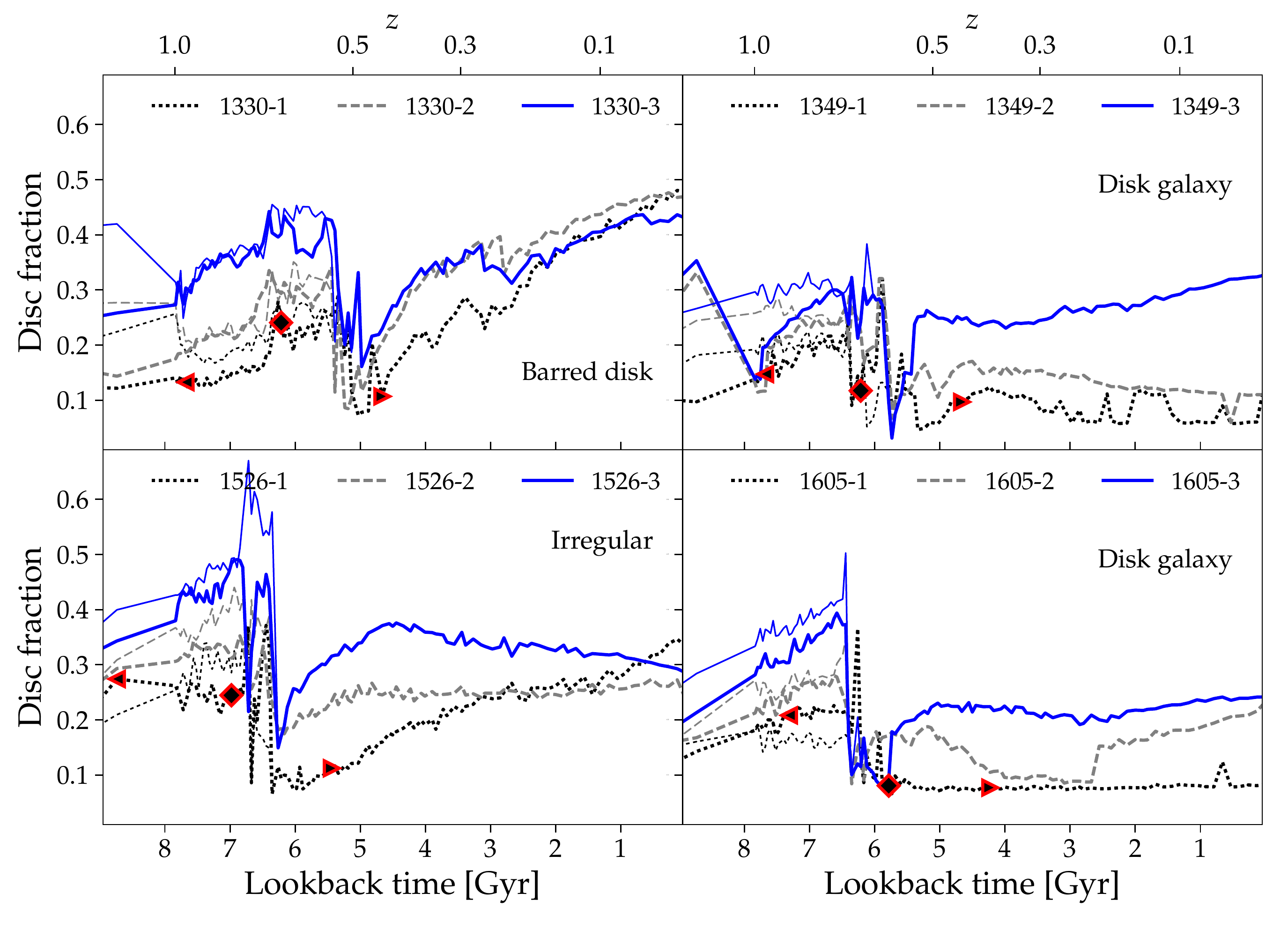}
\caption{The fraction of stellar mass classified as being in the disk according to our kinematical disk--bulge decomposition (\emph{see text for details}). In all simulations the fraction of disc stars is reduced immediately after the major merger, but several of our galaxies have remarkable mass fractions of disk stars larger than $0.3$ at $z=0$. This shows 1) that a galaxy with a dominating spherical component can regrow its disc after a merger and 2) that a merger remnant might very well be a disk galaxy. The lookback times indicated by $\triangleleft$, $\diamond $ and $\triangleright$ mark when the merger starts, peaks and ends, respectively. The peak is determined based on the peak black hole accretion rates of the simulations with a `zoom factor' of 1, and the start and end is 1.5 Gyr before and after this peak, respectively. The secondary galaxy participating in each merger is shown with a thin line. In each panel the classification of the high-resolution simulation revealed by the visual classification in Figure~\ref{PlotDensityMapAtRedshift0} is noted (the same classification is shown in Figure~\ref{AllDiskAge}-\ref{AllSSFR}). The \emph{solid}, \emph{dashed} and \emph{dotted lines} show simulations with zoom factor 3, 2 and 1 (i.e. high, intermediate and low resolution), respectively.}
\label{AllDoverT}
\end{figure*}

The parameters we compare are the disk fraction, the bulge effective radius ($R_\text{eff}$, which is the half-mass-radius of the bulge) and $R_\text{d}$. The disk fractions for the merger remnants are relatively similar to those from the Auriga sample. This supports the claim that a major merger at $z=0.5-1$ does not necessarily exclude the possibility that a galaxy can have a stellar disk at $z=0$. Interestingly, the bulge effective radius for all the merger remnants is in the upper 20\% percentile for the Auriga galaxies, implying that a major merger inflates the physical bulge size. The scale radii of the stellar disks of the merger remnants are very similar to the Auriga galaxies.

\subsection{The mass fraction of disk stars from a kinematical disc--bulge decomposition}

To introduce a more physical estimator of a galaxy's morphology we calculate the fraction of stars in the disk and bulge based on a kinematical disk--bulge decomposition procedure described in \citet{2014MNRAS.437.1750M}. The idea behind this method is to compare the angular momentum of each star around the galaxy's rotation axis to the angular momentum of a circular orbit. First, we diagonalise the moment of inertia tensor of the stellar distribution and determine the eigenvector with the largest absolute value of the angular momentum. The coordinate system is then rotated so the $z$-axis is along this eigenvector. We define a circularity parameter based on the angular momentum of a circular orbit \citep[as in][]{2009MNRAS.396..696S} at a given distance from the center of a galaxy,
\begin{align}
\epsilon \equiv \frac{j_z}{\sqrt{G M(<r) r}}.
\end{align}
Here $j_z$ is a stellar population particle's specific angular momentum in the $z$-direction, $G$ is the gravitational constant, $r$ is the radius and $M(<r)$ is the cumulative mass distribution. We then define all stars with $\epsilon >0.7$ to be disk stars, and the remaining stars to be bulge stars. With this definition stars in a counter-rotating disk are not counted as disk stars. We refer to the fraction of stellar mass in the disk as the `disk fraction'. In our calculation we only study galaxies within 10\% of $R_{200}$ of the halo containing each galaxy.

Figure~\ref{AllDoverT} shows the mass fraction of stars belonging to the disc as a function of the lookback time. Before the merger the most massive progenitor (defined as the galaxy with the largest $M_*$ at $z=0.93$) is shown in \emph{thick blue} and the other progenitor is shown in \emph{thin blue}. All our simulations show that a major merger has a destructive influence on the disk, but a disc fraction of around $\gtrsim 0.25$ is again re-established after the merger for all the high-resolution runs.

For the 1330-3 and 1526-3 galaxies, a disc is regrown within 1-2 Gyr. For the latter simulation the growth of the disc then stops\footnote{This happens because star formation starts to mainly occur outside the established disk of the 1526-3 galaxy at a lookback time of $\simeq 4$ Gyr. Before this time around 80-90\% of the stellar mass is formed in the disk, but at a lookback time of 3.5 Gyr it has declined to 50\%. At $z=0$ only 1\% of the star formation occurs in the disk.} , and the disk fraction declines. The declining disk fraction at low lookback time is consistent with Figure~\ref{PlotDensityMapAtRedshift0}, where a visual disk is absent. For the former simulation a dip in the disk fraction occur at a lookback time of 3 Gyr, and the disk then slowly starts growing again until $z=0$. Based on these simulations we see that merger remnants -- with a weakened disc after a merger -- can regrow a stellar disk (with disk fractions up to 0.45) before $z=0$. For 1605-3, the $z=0$ galaxy has a disk fraction comparable to the value at the end of the merger (see time marked with a $\triangleright$ symbol). This constant disk fraction could either be a result of very little star formation occurring in the galaxy, or it could be a result of a self-regulated galaxy that forms stars at the same rate as stellar mass is removed from the disk.

\begin{figure*}
\centering
\includegraphics[width = 0.8 \textwidth]{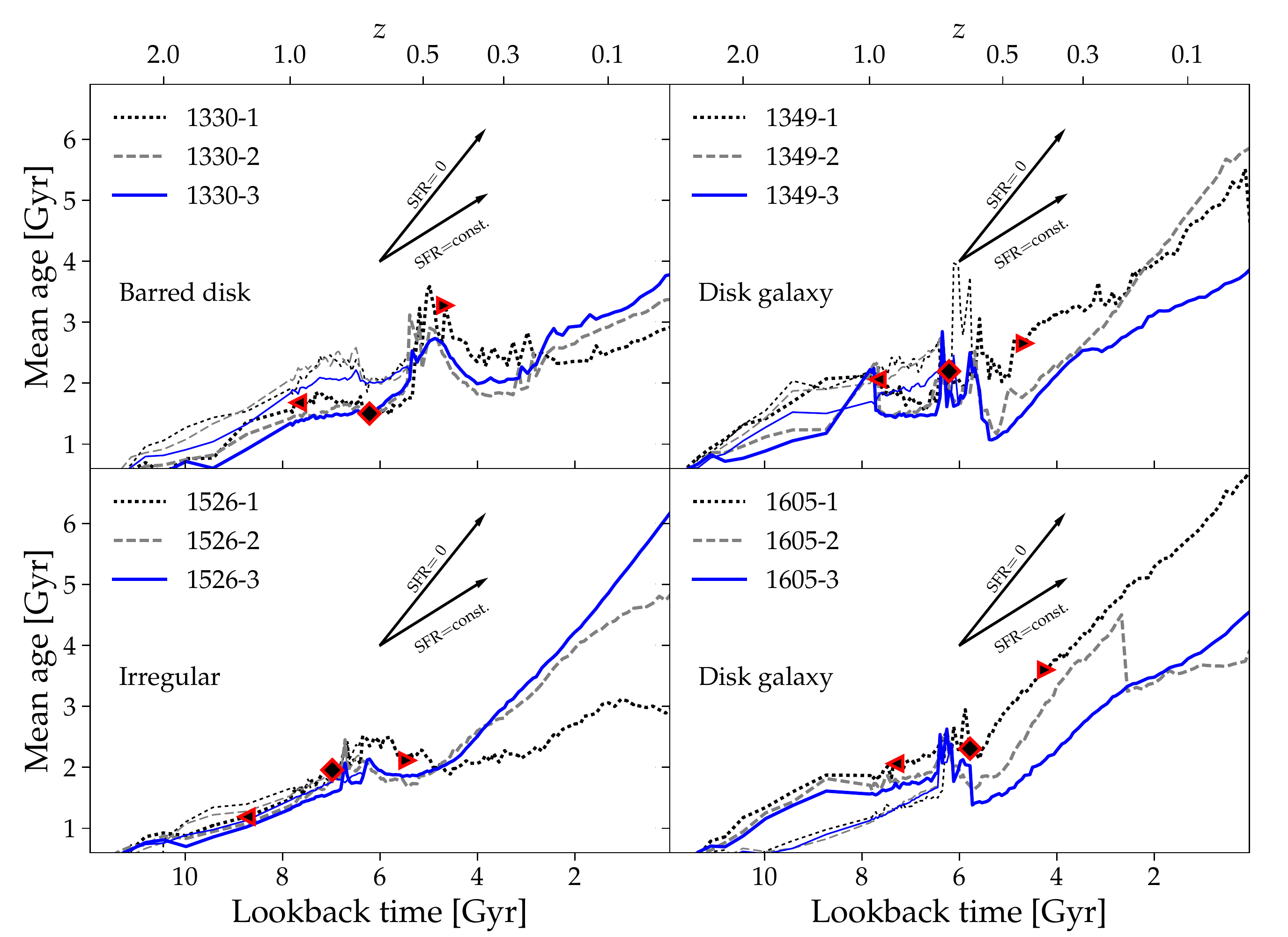}
\caption{The age of the main progenitor's disk as a function of lookback time. For each snapshot we identify the disk stars with $\epsilon>0.7$, and measure their mean age at the given cosmic epoch. The two arrows show tracks of 1) a galaxy without ongoing star formation in the disk ($\text{SFR}=0$) containing only old stars and 2) a constant SFR in the disk throughout the galaxy's lifetime. One of our high-resolution simulations (1526-3) follows a track with no active star formation in the disk, and the three others show signs of recently formed disk stars.}
\label{AllDiskAge}
\centering
\includegraphics[width = 0.8 \textwidth]{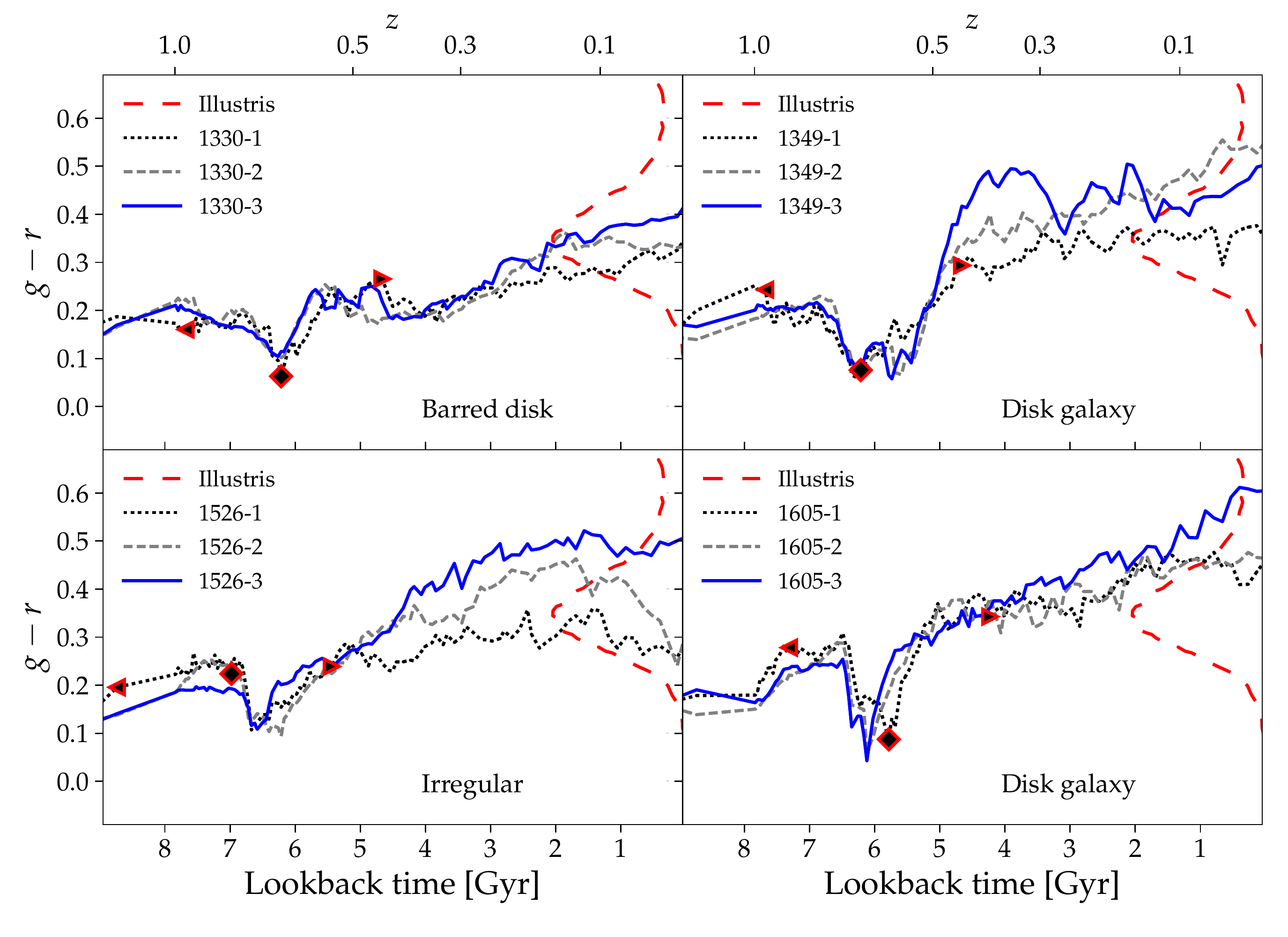}
\caption{The evolution of the $g-r$ SDSS colour for the simulated galaxies. In the time-interval where the mergers occur ($0.5<z<1$), the colour reaches a minimum in all the galaxies. This minimum is caused by the many young stars formed in the merger. In the post-merger evolution the galaxies become redder. The 1330-3 galaxy ends up with a $z=0$ colour similar to the mean colour of galaxies in Illustris, implying that merger remnants are not necessarily redder than normal star-forming galaxies.}
\label{AllColors_gr}
\end{figure*}

Only for the 1330-X simulations we see a similar evolution of the disk fraction for the three resolution levels, implying that the disk fraction is converged in this case. In the other simulations (1349-X, 1526-X and 1605-X) increasing the resolution increases the strength of the disk significantly, and we have thus not established that the evolution of the disk fraction is converged in the high-resolution runs. The resolution at which a simulation converges depends on the details of the evolution history of the galaxy (the same is seen in the Auriga simulation, see Fig. 23 in \citealt{2016arXiv161001159G}).

The disk properties of major merger remnants have also been studied in a set of idealised simulations based on semi-analytic merger models (the method is described in \citealt{2014MNRAS.437.1027M}). \citet{2015MNRAS.451.2968F} found this setup to reduce the bulge formation efficiency compared to semi-analytical galaxy formation models \citep{2007MNRAS.375....2D,2008MNRAS.391..481S,2009MNRAS.399..827L,2014MNRAS.444..942P}, and this setup also makes it possible to examine the role of a transfer of hot gas to the cold disk in the merger process \citep{2015MNRAS.452.2984K}. \citet{2015MNRAS.452.4347K} reported the disk fraction of two major merger remnants from this setup to be 0.22 and 0.33 (for galaxies 18989 and 215240 in their sample, respectively), which is in good agreement with our disc fractions. The results from \citet{2015MNRAS.452.4347K} support the conclusion that merger remnants can have star-forming disks.

\begin{figure*}
\centering
\includegraphics[width = 0.98 \textwidth]{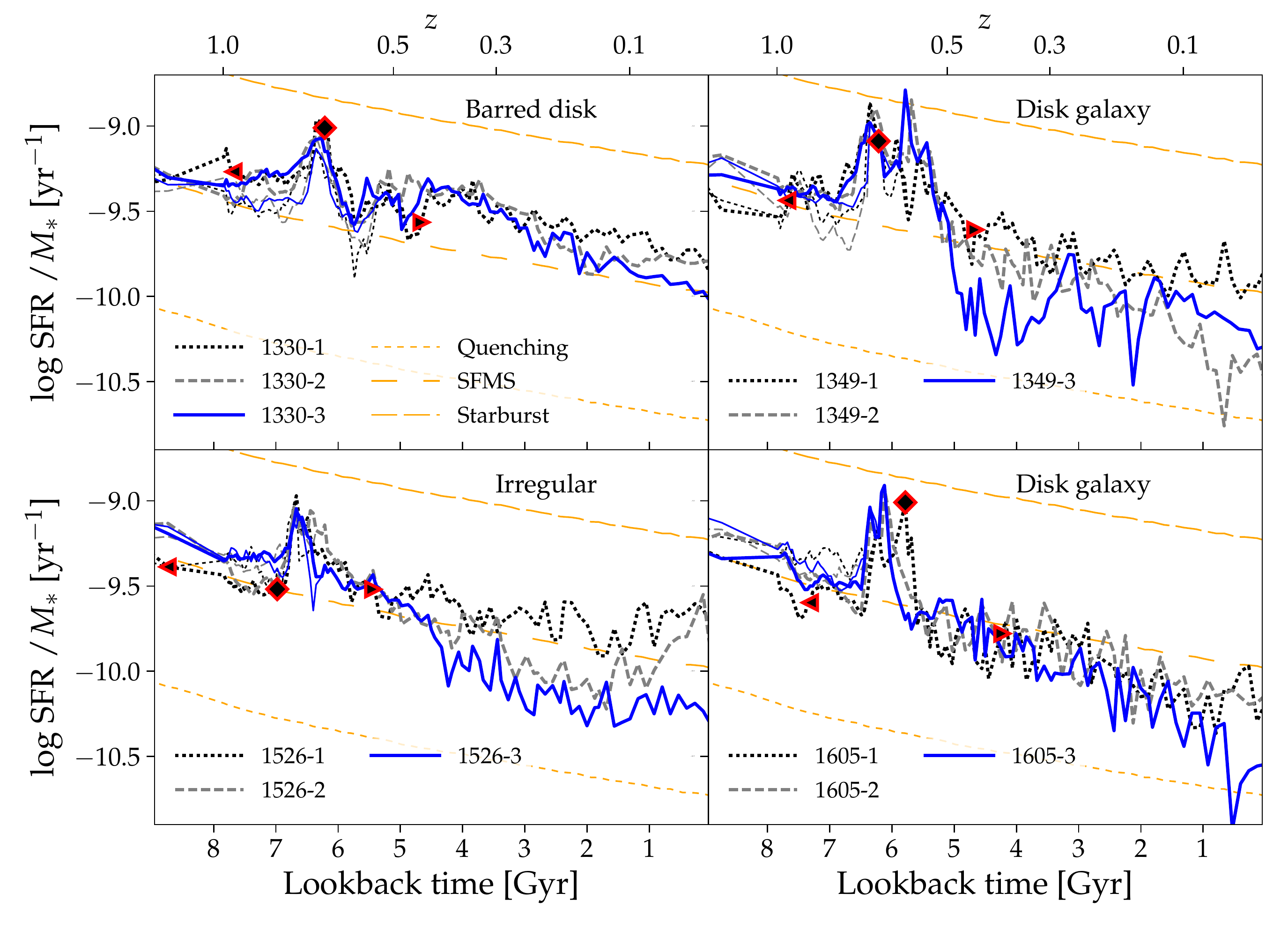}
\caption{The specific star formation rate, SFR$/M_*$, as a function of time. We compare our galaxies to the star formation main sequence from the Illustris simulation \citep{2015MNRAS.447.3548S}, and based on this we define a starburst and quenching threshold as being 0.75 dex above and below, respectively. At $z=0$ the 1605-3 galaxy is close to being quenched. The same is the case for 1349-3 in the first couple of Gyr after the merger, but at $z=0$ the specific SFR indicates that it is a normal star-forming galaxy. The two other high-resolution simulations have a specific SFR indicating that they are normal galaxies belonging to the star formation main sequence. The behaviour challenges the simple galaxy evolution picture, where the feedback followed by a merger-induced starburst automatically quenches galaxies.}
\label{AllSSFR}
\end{figure*}

\subsection{Age of the stellar disk and self-regulation}

The merger simulations exhibit very different evolution-histories of the disk fraction. In the following we will see how these trends can be related to whether active star formation occurs in the disk. As an indicator of this, we first study the evolution of the \emph{mean age} of the disk. At each snapshot we select the disk stars (with $\epsilon>0.7$ and radius smaller than $0.1R_{200}$) in the main progenitor of the $z=0$ merger remnants and calculate their mean age at the given epoch. The resulting relation is shown in Figure~\ref{AllDiskAge}. Also shown are arrows indicating two tracks: the steeper track shows a scenario where the disk is made up of a single-age stellar population and no active star formation. The age of the disk is therefore equal to the age of the single-age stellar population. The less steep track shows a galaxy that has formed disk stars at a constant SFR throughout its lifetime, so the average age of the disk is half the age of the galaxy. We note that these tracks ignore mass loss from the stellar population, and also ignore the transfer of stars from disk to the halo or bulge (we note that all these effects are self-consistently included in our simulations, they are only ignored in the tracks shown by the arrows). If the SFR increases with time we expect a shallower track compared with the shown tracks.

We focus on the post-merger evolution of the galaxies. In the last 4 Gyr of the simulation, the 1526-3 galaxy evolves very similarly to the $\text{SFR}=0$ track, i.e. no active star formation occurs in the disk. The declining disk fraction (noticed in Figure~\ref{AllDoverT}) and the lack of a visible stellar disk (Figure~\ref{PlotDensityMapAtRedshift0}) are thus effects of very few disk stars being formed at low redshift. At $z<0.5$, the two spiral galaxies (1349-3 and 1605-3) evolve similarly to the track expected for a constant SFR, implying that new stars are steadily formed in the disks. The bar/spiral galaxy (1330-3) shows signs of recent star formation activity beyond what is expected for a constant SFR. Detailed inspection of individual snapshots reveals that several mergers and interaction events occur for this galaxy (see Appendix~\ref{time-series} for a more detailed analysis), which is likely the reason for these bursts of star formation. These events also explain why this galaxy has a larger disk fraction than the other galaxies.
In the two cases where an ordered disk is present at $z=0$ we have a constant SFR and stellar mass is continuously added to the disk.

\subsection{Galaxy colours}

Observed galaxies at $z\simeq 0.2$ have a bimodial distribution of the $g-r$ colour, with \emph{red galaxies} having $g-r\gtrsim 0.7$ and \emph{blue galaxies} having $g-r\lesssim 0.7$ \citep{2005ApJ...629..143B}. For the galaxies in the Illustris simulation, \citet{2014MNRAS.444.1518V} also found a bimodial distribution, but it was noticed that the red peak is less pronounced than in observations, and the separation between red and blue galaxies occurs at a slightly too blue colour of $g-r\simeq 0.6$. \citet{2014MNRAS.444.1518V} noted that these differences are probably due to the halo quenching mechanism (AGN feedback) in Illustris not being efficient enough in massive galaxies.

\begin{figure*}
\centering
\includegraphics[width = 0.98 \textwidth]{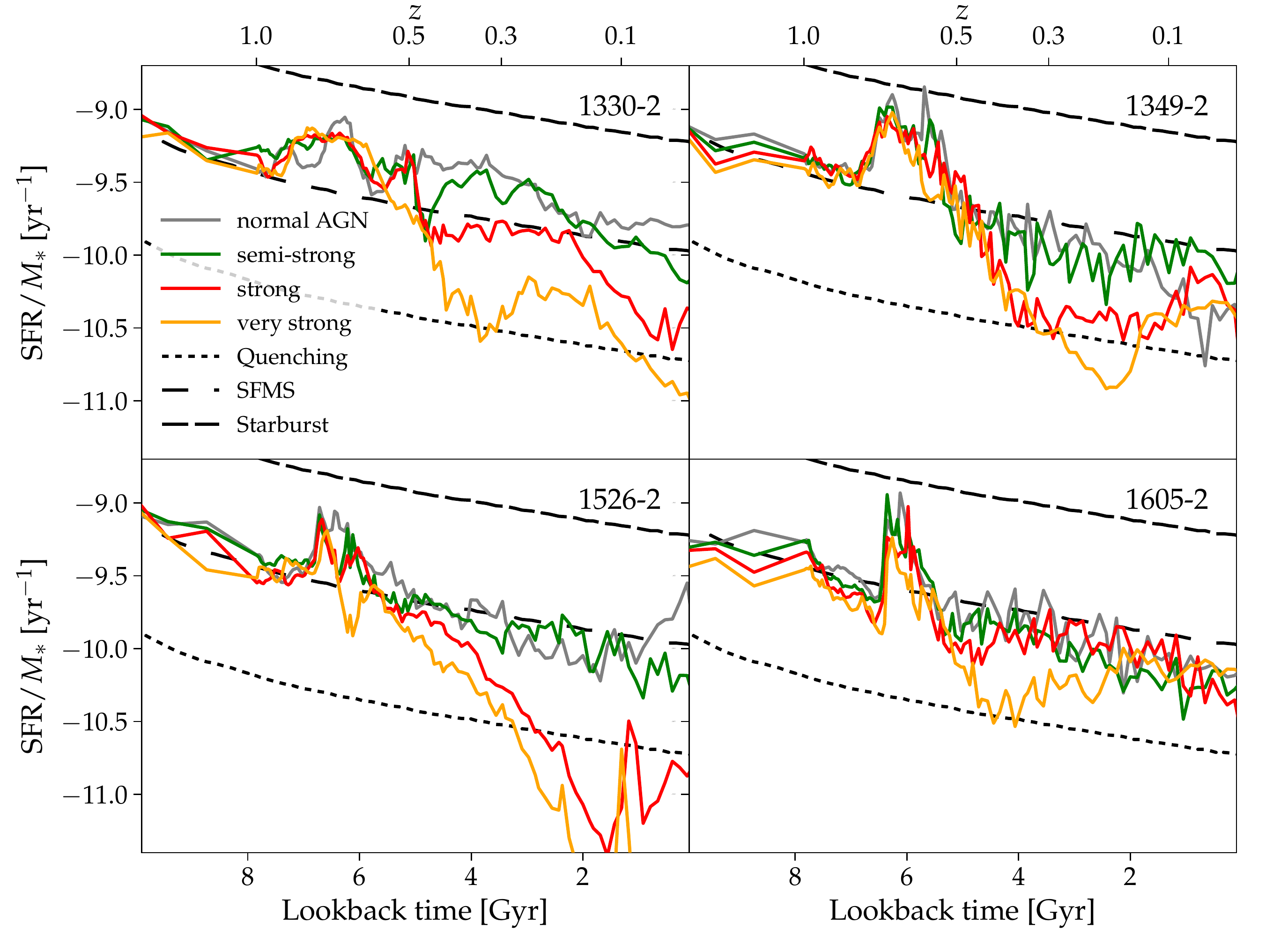}
\caption{The time-evolution of the specific star formation rate for runs with different strengths of the AGN feedback. When the AGN feedback is made stronger (compared to our fiducial level) several of our galaxies reach SSFRs lower than indicated by the quenching threshold (\emph{black--dashed curve}). The figure shows that quenching of merger remnants can be obtained within the framework of the Auriga galaxy formation model.}
\label{AllSSFR_BHVar}
\end{figure*}

In Figure~\ref{AllColors_gr}, we plot the $g-r$ colour for our galaxies. Also shown is the colour distribution for the galaxies in Illustris with $M_*>10^9\msun$ (\emph{dashed line}; instead of lookback time the $x$-axis shows the probability density). The colour is calculated entirely based on synthesis modeling of the stellar populations in the galaxies. During the major merger, where the starburst occurs, all galaxies get bluer in terms of their $g-r$ colour. The typical minimum value is $g-r\simeq 0.1$. After the merger-induced star formation peaks the galaxies settle into a more moderate star formation mode, and the galaxies become redder as time passes. At $z=0$ there are merger remnants having both blue and red $g-r$ colours. The 1330-X simulations for example all end up near the blue $g-r$ peak of the Illustris galaxies. This reflects the fact that the merger remnants remain actively star forming after the merger.

Contrary to these 1330-X galaxies are runs 1349-3, 1526-3 and 1605-3, which are redder with $g-r>0.5$ at $z=0$. This shows that a major merger at $0.5<z<1$ most often leaves a $z=0$ merger remnant with a colour redder than the mean of the galaxy population.

\subsection{Starburst and quenching phases diagnosed by the specific SFR} \label{SecSSFRQuenching}

An essential characterisation of a galaxy is whether it is quenched, normal or a starburst. We will here quantify this by examining the specific SFR, $\text{SSFR}\equiv \text{SFR}/M_*$, of the galaxies, see Figure~\ref{AllSSFR}. This quantity has been computed based on the star formation rate and stellar mass at each outputted simulation snapshot. We compare the galaxies with the SSFR of main sequence galaxies with $M_*=10^{10}\msun$ at $z=0$ from Illustris (taken from Fig. 2 in \citealt{2015MNRAS.447.3548S}). This corresponds to the normalisation of the star formation main sequence, because the simulated main sequence relation has a slope of d$\log$(SFR)/d$\log M_*\simeq 1$. Also shown is the threshold above which galaxies are characterized as starbursts, and the threshold below which a galaxy is considered quenched. The thresholds are selected to be 0.75 dex away from the mean of the main sequence (i.e. 3$\sigma$ away from the main sequence, given that the scatter in Illustris is around 0.25 dex).

First, it is seen that all the galaxies experience a merger-induced SSFR peak at a lookback time between 5 and 7 Gyr. Only one of the high-resolution galaxies (1349-3) gets a SSFR larger than the starburst threshold. Note, however, that we have a snapshot-spacing of 54 Myr during the merger. If we instead of calculating the instantaneous SFR in each snapshot estimate the averaged SFR in 28 Myr intervals (this can be done based on the formation times and initial stellar masses of the stellar population particles in the merger remnant, as done in Fig. 1 in \citealt{2016arXiv160408205S}) we find that the peak SSFR is increased above the starburst threshold for both 1349-3 and 1605-3. The 1330-3 and 1526-3 simulations never get a SSFR larger than the starburst-threshold, which is not surprising, since dense starbursting gas with short gas consumption timescales never arises in these simulations (unlike for the 1349-3 and 1605-3 simulations), as pointed out in \citet{2016arXiv160408205S}.

The merger remnants are not necessarily quenched. The 1330-X and 1526-X runs are for example characterised as \emph{main sequence galaxies} throughout their evolution. Such a scenario is consistent with \citet{2014MNRAS.443L..49P}, which by combining observed galaxies with insight from numerical simulations \citep[for a detailed method description, see ][]{2009A&A...507.1313H} found mergers at $z=0.4-0.75$ to account for a significant fraction of the star formation going on along the main sequence.

\citet{2015MNRAS.454.1840K} pointed out that Illustris has too few passive galaxies compared to observed samples. It is therefore possible that the lack of quenching for our merger remnants is a manifestation of a \emph{diversity problem} for galaxies formed according to our galaxy formation model. Another possible interpretation is that major mergers (and the associated black hole feedback) not necessarily need to quench galaxies, and that some of the star-forming galaxies in our local universe could have undergone a major merger at $z<1$. In Section~\ref{Discussion}, we further discuss these possibilities.

Even though 1349-3 never formally quenches, it has a decreased SSFR a few Gyr after the merger (at a lookback time of 5 to 3.5 Gyr). This is followed by a short time where it is in the middle of the star formation main sequence. This suggests that a galaxy can go from being (close to) quenched, and evolve back to being star-forming. At the same time as it goes into a more star-forming regime, the disk fraction also increases as revealed by Figure~\ref{AllDoverT}.

\subsection{Simulations with strong AGN feedback}\label{StrongAGN}

A remarkable result of Section~\ref{SecSSFRQuenching} is that merger remnants in our simulations rarely quench. In this section we will study how robust this result is to changes in the strength of the AGN feedback. In Figure~\ref{AllSSFR_BHVar} we show the evolution of the specific star formation rate for the various runs with gradually increased strength of the AGN feedback (as described in Section~\ref{SetupStrongAGN}). Common for all runs is that the pre-merger stage ($z>1$) is not strongly affected by the strength of the AGN feedback, and furthermore, the pre-merger galaxies are classified as \emph{star-forming galaxies}. In the post-merger galaxies at $z<0.5$ the situation is very different. Here the galaxies with \emph{very strong} AGN feedback in some cases have a 10 times lower SSFR than for the runs with \emph{normal} AGN feedback. We also see that the runs with \emph{very strong} AGN feedback in all the simulations are able to quench a galaxy a fraction of the time after the merger. 1330-2 and 1526-2 are for example quenched at $z=0$, and 1605-2 and 1349-2 go through a quenched phase (at lookback times of 2.5 Gyr and 4 Gyr, respectively) after which they rejuvenate. We regard it as important to have established that our galaxy formation model can quench merger remnants and in some cases also rejuvenate them. It is, however, still worthwhile stressing that a major merger does not necessarily imply a galaxy being quenched for the rest of its lifetime. Whether a merger remnant is quenched depends on the details of the AGN feedback model, and even when the AGN feedback is made sufficiently strong to quench, rejuvenation may still occur.

\section{Discussion} \label{Discussion}

The $z=0$ properties of our high-resolution (with \emph{normal} AGN feedback) major merger remnants are summarised in Table~\ref{MergerRemnants}. The different simulations of major mergers at $z=0.5-1$ give very diverse $z=0$ galaxies. At one end of the spectrum is the blue star-forming barred galaxy (1330-3) with a disk that accounts for 43\% of the stellar mass in the galaxy. The other extreme is 1605-3 which is close to being quenched and has a galaxy colour characterising it as a \emph{green-valley} galaxy. It furthermore has a low disk fraction of 24\%, but despite of this low value, the disk is still clearly visible in the mock-image of the galaxy. In-between these two extreme galaxies are the other galaxies (1349-3 and 1526-3), which have \emph{blue-green} colours, and are slightly less star-forming than the mean of the star formation main sequence. Their morphologies, as revealed by the optical images, show an ordered disk and a bulge surrounded by an irregular star-forming structure, respectively. For the runs with \emph{very strong} blackhole feedback, quenching occured in all galaxies, but two of the galaxies were also rejuvenated before $z=0$. Within the framework of our galaxy formation model, quenching of merger remnants is therefore a possibility, even though our model tends to prefer the formation of a star-forming disk.

The idea that major mergers might evolve into star-forming galaxies is very different to the standard picture based on idealised simulations, where the AGN feedback associated with the merger quenches the remnant  \citep{2005Natur.433..604D,2005MNRAS.361..776S,2006ApJS..163....1H}. The reason for this fundamentally different behaviour is that our cosmological simulation setup allows gas accretion in the post-merger stage, unlike previous idealised merger simulations.

\begin{table}
\centering
\begin{tabular}{c|cccl}
\hline\hline
Name  & $\log \frac{\text{SSFR}}{\text{SSFR}_\text{SFMS}} $ &$g-r$ & Disk fraction & Visual image\\
\hline
1330-3 & $-0.06$ dex & 0.42  & 0.43 & Barred disk\\
1349-3 & $-0.31$ dex & 0.50 & 0.33 & Disk\\
1526-3 & $-0.34$ dex & 0.51 & 0.29 & Irregular\\
1605-3 & $-0.56$ dex & 0.60 & 0.24 & Disk\\
\hline\hline
\end{tabular}
  \caption{A summary of properties for the $z=0$ major merger remnants from our high-resolution simulations (with \emph{normal} AGN feedback). The columns show the 1) simulation name 2) offset from the star formation main sequence 3) the $g-r$ colour 4) the disk fraction based on a kinematical disk-bulge composition and 5) the visual appearance of the galaxies.}
\label{MergerRemnants}
\end{table}

\subsection{The role of the AGN feedback model}

In this paper we have only studied the role of black hole feedback by varying the parameters of our particular model. Of course, as the physics of blackhole feedback is poorly understood, other models might lead to different results. For example, \citet{2016arXiv160702507P} recently also found a major merger remnant in a cosmological simulation to be quenched. This is in good agreement with our runs with \emph{strong} or \emph{very strong} AGN feedback. In the simulations of \citet{2016arXiv160702507P} quenching occured around 0.5 Gyr after the merger. This is significantly shorter than the $\simeq3$ Gyr quenching timescale we see for our runs with \emph{very strong} AGN feedback in Figure~\ref{AllSSFR_BHVar}.

Observations of quenching timescales could hence provide useful constraints of AGN quenching models. \citet{2014MNRAS.440..889S} for example observed the quenching timescale for green valley galaxies to depend strongly on the galaxy morphology, with early-type galaxies having significantly shorter quenching timescales than late-types. Such observations can potentially constrain numerical AGN models, but a set of simulations with different AGN feedback implementations would be required for such a comparison, which is beyond the scope of this paper.

\citet{2016arXiv160904398H} recently constrained the mass-dependence of the quenching timescale from the evolution of the stellar mass function, the SFR--$M_*$ relation and the quiescent fraction. At $M_* = 10^{10.5}-10^{10.8} \msun $, it typically takes $2.5-4.0$ Gyr for a galaxy to migrate from being a star-forming galaxy to a quenched galaxy. This evolution timescale is well consistent with the quenching timescale for the subset of our runs where quenching occurs.

An approach to black hole feedback modeling that might influence the evolution of merger remnants is the model of \citet{2016arXiv160703486W}, where a kinetic feedback mode is conjectured for massive galaxies with low black hole accretion rates. This mode is introduced to avoid rejuvenation of massive galaxies. The kinetic mode preferentially kicks in for galaxies with $M_{200}\gtrsim 10^{12}\msun$. This threshold is similar to the $z=0$ halo mass of our merger remnants, suggesting that at most a subset of our galaxies would be significantly affected. If we, however, had considered mergers of more massive galaxies, rejuvenation would most likely be prevented by the mechanism proposed by \citet{2016arXiv160703486W}.

\subsection{Relating to the observational quenching framework}

The finding that several of our simulated merger remnants are star-forming and have stellar disks challenges the idea that feedback processes related to a merger-induced starburst always lead to quenching.
\citet{2013ApJ...776...63F} suggests quenching to involve two processes: the star formation in the galaxy has to be shut down in the central parts, and the fueling of gas from the surroundings has to be stopped. This hypothesis is put forward based on observations showing that a dense central bulge is by itself not determining whether massive galaxies are quenched. The observations of \citet{2015MNRAS.448..237W} find most quenched galaxies to have either dense bulges or massive dark matter halos (or both), which thus supports the existence of two quenching mechanisms. In their nomenclature, the reason why our galaxies remain star-forming could be that the dark matter halo masses are not massive enough for halo quenching to occur, even though strong feedback processes are responsible for decreasing the SFR.

A result from our analysis is that a quenched galaxy can transition into a star-forming phase lasting for a few hundreds of Myr (as we saw for 1349-3 in Figure~\ref{AllSSFR}), and then either remain star-forming for the rest of its lifetime or quench again. Such a scenario is consistent with the evolutionary path outlined by \citet{2015MNRAS.450.2327Z} and \citet{2016MNRAS.457.2790T}, where cosmic streams can ignite star formation in galaxies. In the FIRE simulation suite \citep{2014MNRAS.445..581H} a similar re-ignition is also seen as a result of stellar feedback \citep{2015arXiv151003869S}, and analytical considerations support the same process \citep{2015arXiv151005650H}. It is thus well motivated by theoretical considerations that a quenched galaxies can evolve back into the star-forming population. Various groups have studied whether filamentary accretion streams can potentially lead to re-ignition-events \citep{2003MNRAS.345..349B,2005MNRAS.363....2K,2015MNRAS.448...59N}, and even though past hydrodynamical methods likely overestimated the importance of such streams \citep{2013MNRAS.429.3353N}, it remains likely that at least a fraction of galaxies get fed through such events.

\subsection{Disk fractions of galaxies from Illustris}

A natural extension of this project would be to see how merger remnants behave in a larger set of simulations. One possibility to study this would be to simply run more simulations of major mergers with a similar setup as presented in this paper. Another possibility is to study the behaviour of galaxies in a large-scale simulation. This would provide much larger samples, but as we have shown a coarser resolution leads to slightly lower disk fractions than in our high-resolution zoom-setup. Despite of this the lessons learned from such large-scale simulations should be quite still useful.

\citet{2016arXiv160909498R} did such an analysis, where they used galaxies from the Illustris simulation to study how the disk fraction depends on the mass fraction of stars formed \emph{ex-situ}. Here the ex-situ fraction is a proxy for the importance of mergers in the assembly history of a galaxy. For $10^{11}\leq M_*/\msun\leq 10^{12}$, the two quantities are anti-correlated implying that mergers destroy disks. Their findings suggest that this anti-correlation is caused by a decreased gas fraction of the merging galaxies. For galaxies with lower stellar masses of $10^{10}\leq M_*/\msun\leq 10^{11}$, the anti-correlation between disk fraction and ex-situ fraction is less pronounced, implying that mergers do not play a significant role in destroying stellar disks; for example several galaxies with ex-situ fractions of $\simeq 0.5$ have significant disk components with mass fractions of $\simeq 0.5$. This is consistent with our finding that merger remnants in this mass-range might evolve into star-forming disk galaxies.

\section{Conclusion} \label{Conclusion}

In this work, we have continued our study of fully self-consistent cosmological high-resolution hydrodynamical simulations of major mergers. A companion paper showed how starbursting gas appears when the mass resolution is made 10-40 times finer than in the Illustris simulation. This suggests a solution to the problem of too few starburst galaxies in the Illustris simulation.

In the present paper, we study how morphological transformations and quenching occur in mergers. Our findings challenge the orthodox idea that a major merger automatically leads to a quenched elliptical. Our main results can be  summarised as follows:

\begin{itemize}

\item Visual inspection, surface-density modeling and a kinematical disk-bulge decomposition reveal that the $z=0$ remnant of a major merger (simulated at our highest resolution) that happened at $z=0.5-1$ can easily have a star-forming disk. Also, the colours of our galaxies classify them as \emph{blue--green} galaxies. This challenges the conventional idea that the black hole feedback associated with a merger-induced starburst quenches galaxies. Idealised simulations have previously shown that a major merger remnant is not necessarily a quenched elliptical, but we here show the same result in a set of cosmological self-consistent galaxy formation simulations.

\item In a set of runs with stronger AGN feedback we show that merger-induced quenching occurs if the feedback is made sufficiently strong. For our runs with the strongest feedback, all galaxies go through a quenched phase, but only two of the four merger remnants stay also quenched up to  $z=0$. This suggests that rejuvenation is important for merger remnants with a halo mass around $10^{12}\msun$. We conclude that \emph{merger-induced quenching is a possibility, but not a requirement} within the framework of our galaxy formation model.

\item In the majority of our simulations with \emph{normal} strength of the AGN feedback, major mergers have a destructive effect on the disk, since the pre-merger galaxies have larger disk fractions than the merger remnant. It does, however, not prevent merger remnants from regrowing a new disk before $z=0$.

\item The bulges of the $z=0$ merger remnants have effective radii which are on average significantly larger than galaxies with a more quiescent accretion history (from the Auriga galaxy simulation suite). The distribution of disk fractions and disk scale lengths are similar between the two simulation samples, again implying that major mergers do not necessarily need to be the main mechanism for driving galaxy transformations.

\end{itemize}

\section*{Acknowledgements}

We thank Chang Hoon Hahn, Lars Hernquist, Rahul Kannan and Vicente Rodriguez-Gomez for useful comments and discussions. We also thank the referee for a constructive report. The authors would also like to thank the Klaus Tschira Foundation. MS thanks the Sapere Aude fellowship program. VS acknowledges the European Research Council through ERC-StG grant EXAGAL-308037, and the SFB-881 `The Milky Way System' of the German Science Foundation.

\def\aj{AJ}
\def\araa{ARA\&A}
\def\apj{ApJ}
\def\apjl{ApJ}
\def\apjs{ApJS}
\def\apss{Ap\&SS}
\def\aap{A\&A}
\def\aapr{A\&A~Rev.}
\def\aaps{A\&AS}
\def\mnras{MNRAS}
\def\nat{Nature}
\def\pasp{PASP}
\def\aplett{Astrophys.~Lett.}
\def\physrep{Physical Reviews}
\def\nar{New A Rev.}

\footnotesize{
\bibliographystyle{mn2e}
\bibliography{ref}
}

\appendix
\clearpage\clearpage

\begin{figure*}
\centering
\includegraphics[width = 0.98 \textwidth]{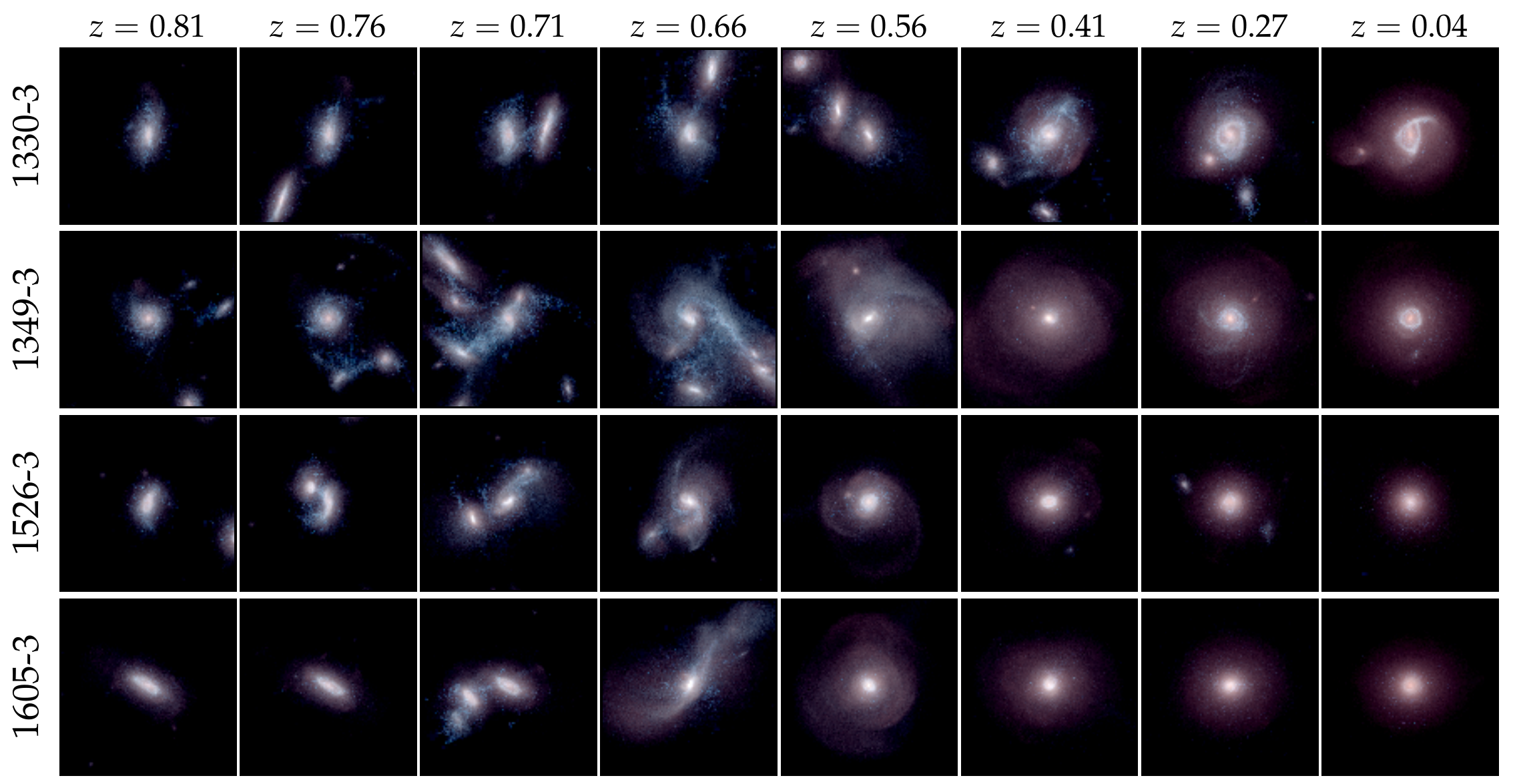}
\caption{The time-evolution of the stellar components of the four high-resolution galaxies. The colour coding is the same as in Figure~\ref{PlotDensityMapAtRedshift0}. Each panel has a width of 80 kpc.}
\label{Fig261_StarsTimeSeries}
\end{figure*}

\section{Evolution history of the mergers}\label{time-series}

The time-evolution of the high-resolution mergers is shown in Figure~\ref{Fig261_StarsTimeSeries}. At $z=0.71$ seven galaxies are visible in the image of 1349-3. This explains why the two most-massive progenitors of this merger remnant approach each other faster than an $E=0$ orbit (see \citealt{2016arXiv160408205S}). In the other mergers only two massive progenitors are present in the pre-merger snapshots.

1330-3 undergoes several minor mergers at low redshift; at $z=0.41$, $z=0.27$ and $z=0.04$ accreted satellites are revealed in the figure. This is accompanied by active star formation in the merger remnant. Figure~\ref{AllColors_gr} for example reveals that this galaxy maintains a blue colour after the merger, and Figure~\ref{AllSSFR} shows that it is slightly above the mean of the star formation main sequence. The gas accretion and tidal disruptions caused by such minor mergers is likely what drives the formation of the bar revealed in the $z=0$ image of the galaxy (see Figure~\ref{PlotDensityMapAtRedshift0}).

\clearpage

\label{lastpage}

\end{document}